\documentclass[amsmath,amssymb,showpacs,reprint,floatfix,aps,prl,superscriptaddress]{revtex4-1}
\usepackage[hidelinks,colorlinks=true,allcolors=blue]{hyperref}
\usepackage{amssymb}
\usepackage{amsmath}
\usepackage{graphicx}
\usepackage{physics}
\usepackage{braket}
\usepackage{color}
\usepackage{bm}
\usepackage[labelfont=bf]{subcaption}
\usepackage[labelfont=bf]{caption}
\DeclareCaptionLabelFormat{noparens}{\textbf{#2}} 
\graphicspath{ {./}{../figures/} }

\newcommand{\kk}{\mathbf{k}}
\renewcommand{\qq}{\mathbf{q}}

\begin{document}
\setlength{\intextsep}{5pt}
\title{
Toward precise simulations of the coupled ultrafast dynamics\\ of electrons and atomic vibrations in materials
}
\author{Xiao Tong}
\affiliation{%
Department of Applied Physics and Materials Science, California Institute of Technology, Pasadena, CA 91125, USA.}

\author{Marco Bernardi}
\email[E-mail: ]{bmarco@caltech.edu}
\affiliation{%
Department of Applied Physics and Materials Science, California Institute of Technology, Pasadena, CA 91125, USA.}

\begin{abstract}

\noindent 
Ultrafast spectroscopies can access the dynamics of electrons and nuclei at short timescales, shedding light on nonequilibrium phenomena in materials. 
However, development of accurate calculations to interpret these experiments has lagged behind as widely adopted simulation schemes are limited to sub-picosecond timescales or employ simplified interactions lacking quantitative accuracy. 
Here we show a precise approach to obtain the time-dependent populations of nonequilibrium electrons and atomic vibrations (phonons) up to tens of picoseconds, with a femtosecond time resolution. 
Combining first-principles electron-phonon and phonon-phonon interactions with a parallel numerical scheme to time-step the coupled electron and phonon Boltzmann equations, our method provides unprecedented microscopic insight into scattering mechanisms in excited materials.  Focusing on graphene as a case study, we demonstrate calculations of ultrafast electron and phonon dynamics, transient optical absorption, structural snapshots and diffuse X-ray scattering. Our first-principles approach paves the way for quantitative atomistic simulations of ultrafast dynamics in materials.

\end{abstract} 
\maketitle
 \vspace{-3pt}
%
%
Time-domain spectroscopies open a window on the dynamics of electrons, phonons and various elementary excitations, 
probing matter on timescales characteristic of the interactions among its constituents. 
In particular, recently developed experimental techniques enable the characterization of the coupled dynamics of electronic and nuclear degrees of freedom with high spatial and time resolutions~\cite{buzzi2018probing, najafi2017super, zurch2017ultrafast, Damascelli2019}, as well as manipulation of ultrafast structural dynamics~\cite{young2010femtosecond, fritz2007ultrafast, reis2006ultrafast, trigo2013fourier, lindenberg2000time, zalden2019femtosecond} and phase transitions~\cite{de2013speed, mankowsky2017ultrafast, mitrano2016possible}.
\\ 
\indent
The vast information encoded in ultrafast spectroscopy signals underscores a critical need for quantitative theoretical tools. 
The latter should ideally be able to predict or interpret ultrafast measurements and shed light on the underlying microscopic processes 
and coupling between different degrees of freedom. 
A common scenario is the coupled dynamics of excited electrons and phonons, which governs a wide range of phenomena such as carrier and lattice equilibration, 
photoemission and the associated linewidths, structural transitions and superconductivity.  
Heuristic approaches such as two-temperature models~\cite{wilson2013two} or kinetic equations with adjustable interactions~\cite{haug2008quantum} are routinely adopted to study 
aspects of nonequilibrium electron and phonon dynamics. As they leave out important atomistic details,  
these models are only occasionally geared toward quantitative predictions~\cite{waldecker2016electron} and are mainly useful as tools to qualitatively interpret experimental results.    
\\
\indent
First-principles computational methods based on density functional theory (DFT) and related techniques have made great strides in modeling electron and phonon interactions and nonequilibrium dynamics~\cite{murray2007phonon, Bernardi-Si, Bernardi-SPP, jhalani2017ultrafast, bernardi2016first, sadasivam2017theory}.
An important approach is real-time time-dependent DFT (rt-TDDFT)~\cite{yabana1996time, castro2004propagators, marques2004time}, which propagates in time the electronic Kohn-Sham equations 
while treating atomic motions using Ehrenfest forces. 
While rt-TDDFT has been widely successful for studying electron dynamics in molecules~\cite{nazeeruddin2005combined, rozzi2013quantum}, 
important open challenges remain, including reaching simulation times longer than $\sim$1 ps, 
treating periodic systems (crystals)~\cite{sottile2005tddft} and better understanding, improving and validating the accuracy of exchange-correlation functionals governing the interactions of electrons and nuclei~\cite{marques2012fundamentals, HeadGordon-2003, Maitra-2012}. 
Non-adiabatic molecular dynamics is also a valuable method to model ultrafast dynamics in molecules~\cite{yang2020simultaneous}, although it is not commonly employed for solids. 
\\
\indent 
A second family of approaches $-$ on which this work hinges $-$ employs perturbation theory to compute the matrix elements for electron and phonon interactions~\cite{Mahan-main}, and combines them with the semiclassical Boltzmann transport equation (BTE)~\cite{jhalani2017ultrafast} or quantum kinetic equations~\cite{sangalli2015complete} to investigate ultrafast electron dynamics. 
Owing to its momentum-space formulation, this framework is ideally suited for studies of crystals and periodic systems. However, in spite of recent progress, propagating in time the BTEs for coupled electrons and phonons while fully taking into account their interactions has remained an elusive goal due to computational cost and the need for complex algorithms and workflows.
\begin{figure*}[t]
\includegraphics[width=0.97\textwidth]{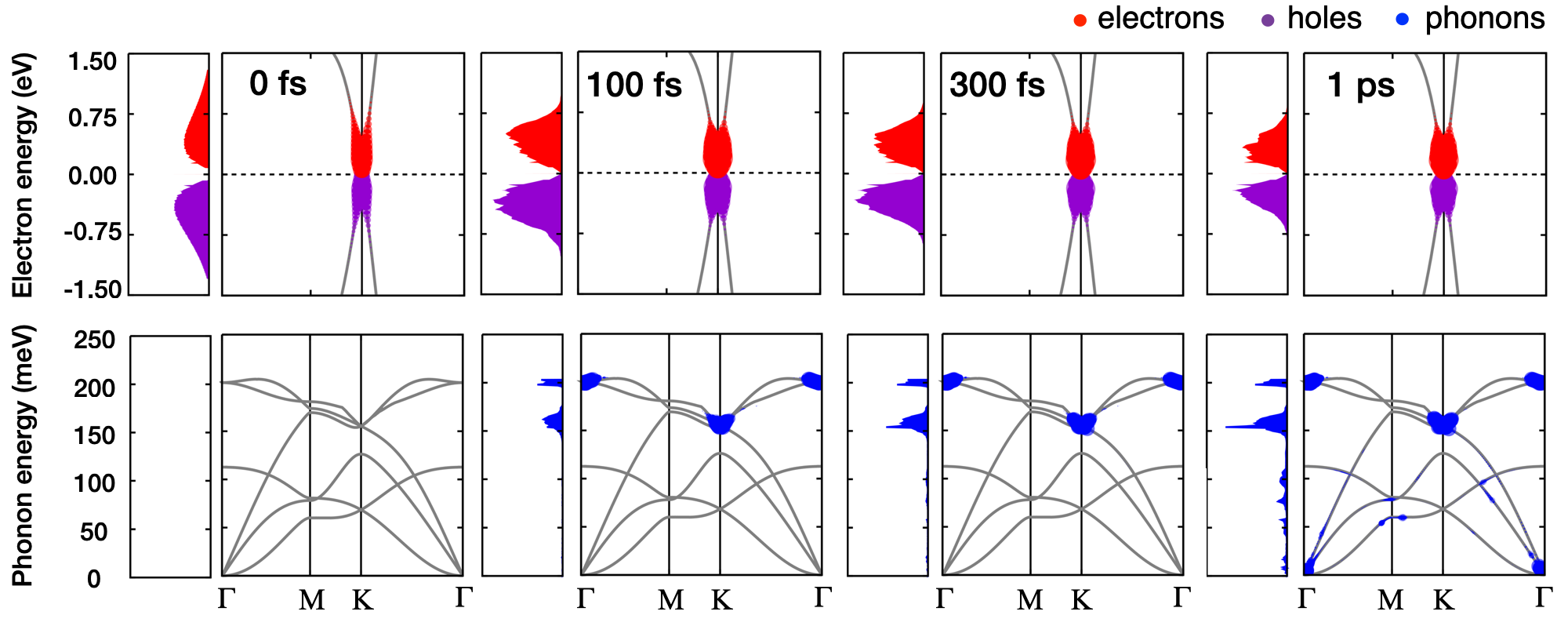}
\caption{\textbf{Simulated carrier and phonon dynamics in graphene.}
Excited electrons (red) and holes (purple) in graphene are initialized in a Fermi-Dirac distribution at 4,000 K electronic temperature at time zero. The subsequent nonequilibrium dynamics is analyzed by mapping the populations $f_{n\kk}(t)$ for electrons and $1 - f_{n\kk}(t)$ for holes on the electronic band structure near the Dirac cone (top) and the phonon populations $N_{\nu\qq}(t)$ on the phonon dispersions (bottom), using a point size proportional to the population values. Shown left of each population panel are the corresponding BZ-averaged carrier or phonon concentrations as a function of energy (in arbitrary units).} 
\label{RTS} 
\end{figure*}
\\
\indent
Here we show explicit simulations of the coupled dynamics of nonequilibrium electrons and phonons for timescales up to tens of picoseconds, 
using accurate electron-phonon (e-ph) and phonon-phonon (ph-ph) interactions validated through transport calculations. 
We time-step the electron and phonon BTEs $-$ a large set of coupled integro-differential equations $-$ using a parallel numerical approach to efficiently compute the relevant collision integrals and obtain time-dependent electron and phonon populations. 
Focusing on graphene, a material with unconventional physical properties and unique promise for ultrafast devices, we investigate the coupled dynamics of excited electrons and phonons, compute spectroscopic signals such as transient absorption and diffuse X-ray scattering, and analyze the dominant scattering channels, identifying the slow rise of flexural phonons as a bottleneck to equilibration.  
Taken together, our work demonstrates a leap forward in simulations of ultrafast dynamics in solids and provides a quantitative tool to predict and interpret time-domain spectroscopies.\\
\begin{figure*}[th]
\centering
\includegraphics[width=0.92\textwidth]{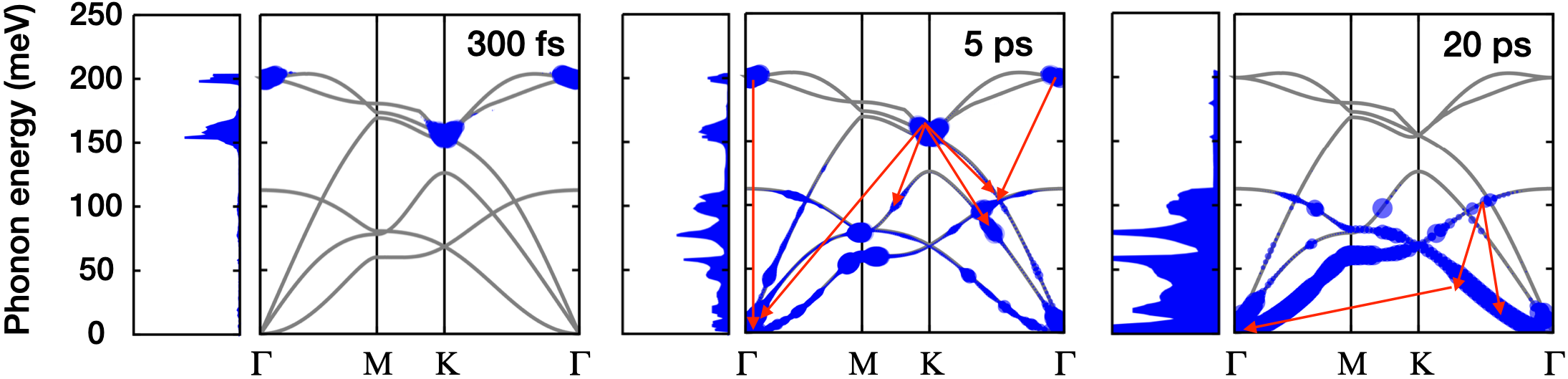}
\caption{\textbf{Phonon scattering channels.} 
Optical phonons at $\Gamma$ and K are rapidly populated in the first 300 fs due to their strong coupling with the excited carriers, but later decay into lower-energy optical and acoustic modes through ph-ph processes. The dominant energy and momentum-conserving phonon decay channels are shown with red arrows at 5 ps and 20 ps delay times after the initial electronic excitation.} \label{channels}
\end{figure*}
\vspace{-20pt}
\section{Results} 
\vspace{-12pt}
\noindent {\bf Numerical approach.} 
We solve the coupled electron and phonon BTEs~\cite{Mahan-nutshell} in a homogeneously excited region of a material  
using a fourth-order Runge-Kutta algorithm. At each time $t$, we propagate the electronic populations $f_{n\kk}(t)$, where $n$ is the band index and $\kk$ the crystal momentum, and the phonon populations $N_{\nu\qq}(t)$, where $\nu$ is the mode index and $\qq$ the phonon wavevector. Using an in-house modified version of our {\sc Perturbo} code~\cite{zhou2020perturbo}, we solve the BTEs as a set of integro-differential equations: 
\begin{align}
&\frac{\partial f_{n\kk}(t)}{\partial t} = \mathcal{I}^{\text{e-ph}}\Big[\big\{f_{n\kk}(t)\big\},\big\{N_{\nu\qq}(t)\big\}\Big] \nonumber \\
&\frac{\partial N_{\nu\qq}(t)}{\partial t} = \mathcal{I}^{\text{ph-e}}\Big[\big\{f_{n\kk}(t)\big\},\big\{N_{\nu\qq}(t)\big\}\Big] +  \mathcal{I}^{\text{ph-ph}}\Big[\big\{N_{\nu\qq}(t)\big\}\Big]\,,
\label{eq1}
\end{align}
where $\mathcal{I}^{\text{e-ph}}$ is the collision integral for e-ph interactions in the electron BTE, while $\mathcal{I}^{\text{ph-e}}$ and $\mathcal{I}^{\text{ph-ph}}$ are, respectively, the collision integrals for phonon-electron (ph-e) and ph-ph interactions in the phonon BTE. 
Detailed expressions for these integrals are given in Methods. 
\\
\indent
The collision integrals depend explicitly on the electron and phonon populations at the current time step, and thus Eq.~(\ref{eq1}) is a system of integro-differential equations, with size $\mathcal{N}_{\text{e}}=\mathcal{N}_b\times \mathcal{N}_\kk$ for electrons and $\mathcal{N}_{\text{ph}}=\mathcal{N}_{\nu}\times\mathcal{N}_{\qq}$ for phonons ($\mathcal{N}_b$ and $\mathcal{N}_{\kk}$ are the number of electronic bands and $\kk$-points, while $\mathcal{N}_{\nu}$ and $\mathcal{N}_{\qq}$ are the number of phonon modes and $\qq$-points). 
In a typical calculation, our scheme involves time-stepping $\sim$$10^6$ coupled integro-differential equations, each with $10^7-10^9$ scattering processes in the collision integrals, a formidable computational challenge addressed with an efficient algorithm combining MPI and {\sc OpenMP} parallelization. 
As memory effects are not included, our BTEs amount to a Markovian dynamics of the density matrix in which the off-diagonal coherences, which are typically short-lived ($<$ 5 fs), are neglected~\cite{haug2008quantum}. Using this scheme, we are able to reach simulation times of up to $\sim$100 ps, with a fs time step.

\noindent {\bf Relaxation of excited carriers.} 
In a first set of simulations, we model photoexcited electron and hole carriers in graphene by choosing appropriate carrier populations at time zero as the initial condition of the BTEs. 
Pump-probe studies on graphene have shown that fast electron-electron interactions rapidly thermalize the excited carriers within 10$-$50 fs~\cite{dawlaty2008measurement, winzer2012impact}. 
Therefore, shortly after excitation, electrons and holes are well described by a hot Fermi-Dirac distribution with a temperature of a few thousand degrees for typical experimental settings~\cite{wang2010ultrafast, sun2012dynamics, gierz2013snapshots}. 
We model the carrier dynamics in graphene after this initial thermalization by setting the electron and hole populations at time zero to Fermi-Dirac distributions at 4,000 K temperature, 
while setting the initial phonon populations to their equilibrium value at 300 K. 
\\
\indent
Figure~\ref{RTS} shows the time evolution of the electron, hole and phonon populations in the first picosecond after the initial excitation and carrier thermalization. 
For representative times, we analyze both the carrier and phonon populations ($f_{n\kk}$ for electrons, $1 - f_{n\kk}$ for holes, and $N_{\nu\qq}$ for phonons) 
along high-symmetry Brillouin zone (BZ) lines, as well as BZ-averaged carrier and phonon concentrations as a function of energy (see Methods).  
In the first 100 fs, the hot electron and hole distributions become narrower in energy as the carriers cool and settle near the Dirac point in about 300 fs. 
This rapid cooling process is achieved by emitting optical phonons through intravalley and intervalley e-ph scattering.  
As a result, the optical phonons generated within 1 ps possess wavevectors close to the BZ $\Gamma$-point for intravalley, and K-point for intervalley scattering. 
\\
\indent
Although most of the excess energy stored in the excited carriers is dissipated during the first 300 fs, 
surprisingly it takes more than 5 ps for the carriers to fully reach equilibrium with negligible population fluctuations. 
This slow and lingering cooling process is due to two factors $-$ one is the slow rise of out-of-plane flexural phonons due to their weak e-ph coupling (see below), 
and the other is the weak anharmonic ph-ph coupling, which allows the optical phonons generated during the rapid carrier cooling to linger for $>10$ ps, 
heating back the carriers through phonon absorption processes.\\

\vspace{5pt}
\noindent {\bf Phonon equilibration.} 
Accessing the 10$-$100 ps time scale characteristic of phonon dynamics is an open challenge for existing first-principles simulations of ultrafast dynamics.  
In Fig.~\ref{channels}, we analyze the long-lived equilibration of optical phonons generated by the excited carriers.  
The optical phonons first decay within $\sim$5 ps to transverse acoustic (TA) and longitudinal acoustic (LA) phonons, 
which then emit lower-energy acoustic modes through ph-ph scattering processes, ultimately equilibrating on a 20$-$100 ps time scale (not shown).
\\
\indent
Our calculations can directly identify the main phonon modes and relaxation pathways from the time-dependent phonon populations (see Fig.~\ref{channels}). 
In the first 300 fs, the excess electronic energy rapidly generates modes with strong e-ph coupling, 
including the A$^{'}_1$ and E$_{2g}$ optical modes, respectively with wavevectors near the K and $\Gamma$ points of the BZ. 
A slower change in phonon populations occurs between 1$-$5 ps, when the optical phonons decay into LA and TA modes with wavevectors halfway between $\Gamma$ and the K or M points of the BZ. Long-wavelength (small-$\qq$) acoustic phonons are also generated in the same time window. 
A second stage of phonon equilibration sets in after $\sim$20 ps, when energy redistribution occurs primarily within the LA and TA branches (see Fig.~\ref{channels}).
\begin{figure}[!t]
\includegraphics[width=\linewidth]{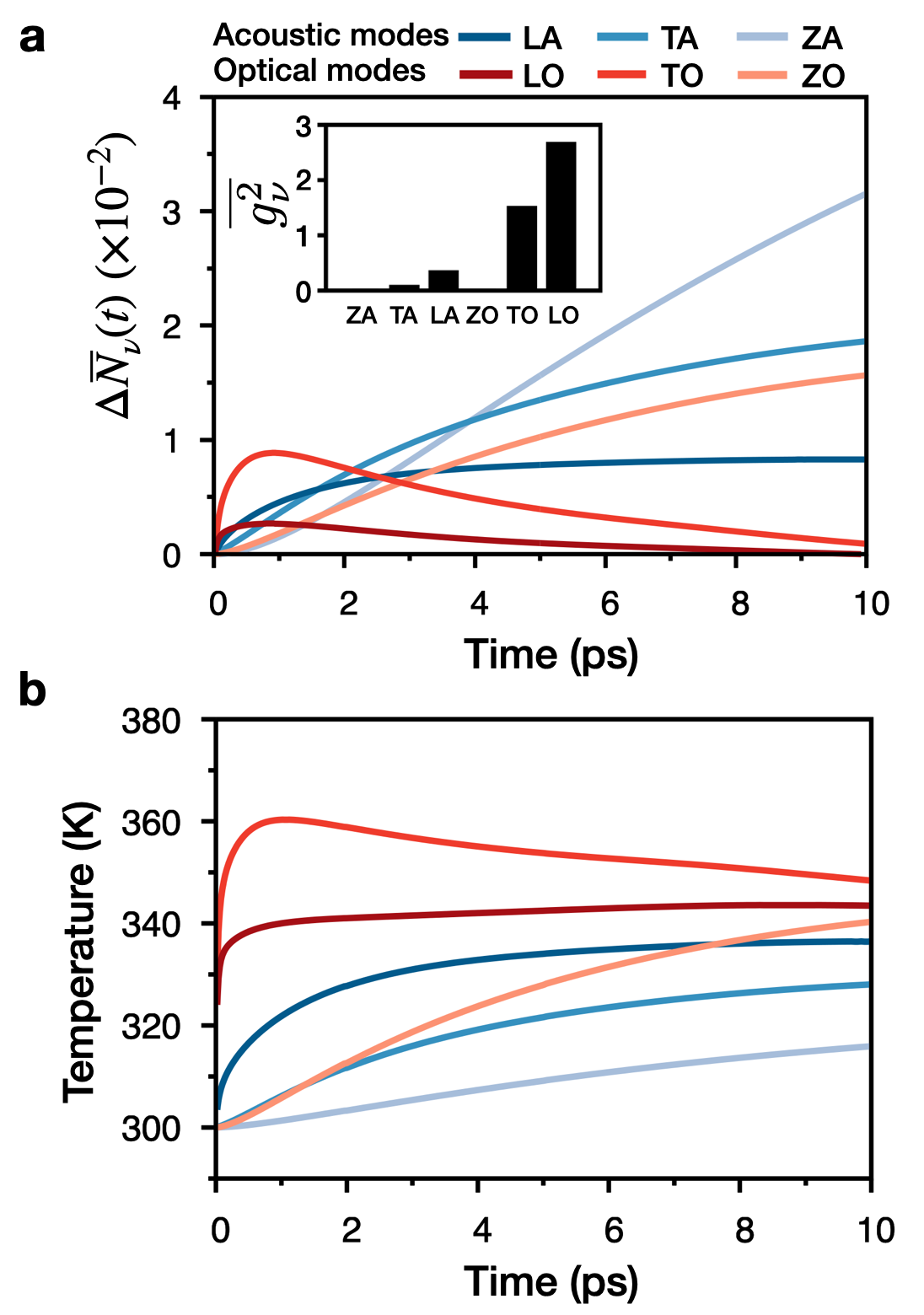}
\caption{\textbf{Mode-resolved phonon populations and effective temperatures.} \textbf{a} Time-dependent excess phonon populations $\Delta \overline{N}_{\nu}(t)$, with e-ph coupling strengths $\overline{g^2_\nu}$ (in arb. units) given in the inset; \textbf{b}, Time-dependent effective temperatures $\overline{T}_\nu$. All quantities are shown separately for each of the six phonon modes in graphene.  
}\label{figure3}
\end{figure}

\indent
We carry out a mode-by-mode analysis of the phonon populations and effective temperatures. 
In Fig.~\ref{figure3}a, we plot for each mode the excess population relative to equilibrium, 
$\Delta \overline{N}_{\nu}(t) = \overline{N}_\nu(t) - \overline{N}_\nu(-\infty)$, where $\overline{N}_\nu(t)$ are BZ-averaged phonon populations for each mode. 
Also shown in Fig.~\ref{figure3}a are the mode-resolved e-ph coupling strengths, $\overline{g^2_\nu}$ (see Methods). 
Due to their strong e-ph coupling, the in-plane longitudinal optical (LO) and transverse optical (TO) modes are rapidly emitted during the initial fast carrier cooling, 
highlighting the key role of e-ph interactions at early times~\cite{Damascelli2019}. 
The LO, TO and LA modes are populated extensively before 1 ps, while the TA and ZO modes are excited more gradually through ph-ph interactions over timescales longer than 1 ps.  
Due to their weak e-ph coupling, the out-of-plane flexural phonon modes (ZA and ZO) are generated at a significantly slower rate than in-plane phonons, 
resulting in a slow rise of their populations in the first 10 ps. The ZA mode, with the weakest e-ph scattering strength, exhibits the slowest rise in population among all modes. 
In turn, the weak e-ph coupling and slow generation of flexural phonons is responsible for a carrier cooling bottleneck: 
As electrons and holes interact with hot ZA and ZO populations, they gain energy from the flexural modes for over 10 ps, well beyond the initial sub-ps carrier cooling.
\\
\indent
Although at short times the phonon distributions are still non-thermal, we find that after 2$-$5 ps all phonon populations are thermal and well approximated by hot Bose-Einstein distributions. 
The mode-resolved effective phonon temperatures are computed as BZ averages of state-dependent temperatures (see Methods) and shown in Fig.~\ref{figure3}b. 
We find that the effective phonon temperatures are strongly mode-dependent throughout the simulation, their trend mirroring the respective excess phonon populations in Fig.~\ref{figure3}a. 
This result shows that a two-temperature model would fail dramatically in graphene, 
both because the phonon temperatures are not well defined before 2 ps and because they are mode-dependent at longer times.  
\\
\indent
Our calculations shed light on the entire equilibration cascade, with timescales spanning several orders of magnitudes,  
from fs for carrier cooling through e-ph interactions to ps for phonon downconversion to equilibration via ph-ph processes over tens of ps. 
The microscopic details of the scattering processes, accessed easily in our approach due to its formulation in momentum space, are out of reach for existing first-principles simulations. 
As shown next, the time-dependent electron and phonon populations further allow us to reliably simulate various ultrafast spectroscopies employed as probes of electron and nuclear dynamics.\\ 
%

\noindent {\bf Pump-probe transient absorption.} 
Time-resolved differential transmission measurements are widely employed to shed light on nonequilibrium carrier dynamics and investigate the timescales associated with electronic processes. 
We build on the calculations shown above to compute transient absorption at optical wavelengths.
\\
\indent
In graphene, at low energies within 1$-$2 eV of the Dirac cone, the differential transmissivity $\Delta T/T_0$ is proportional to the transient carrier populations; 
in a single particle picture and neglecting the state-dependence of the optical transition dipoles~\cite{breusing2011ultrafast}, 
\begin{equation}
\frac{\Delta T(E)}{T_0} \approx -a_0\bigg[ \Delta \overline{f}_\mathrm{e} \bigg( \frac{E}{2} \bigg) + \Delta \overline{f}_\mathrm{h} \bigg(\!-\frac{E}{2}\bigg)\bigg]
\label{eq2}
\end{equation}
where $E$ is carrier energy, $a_0$ is the absorption coefficient of single-layer graphene, $\overline{f}_{\mathrm{e,h}}$ are BZ-averaged electron or hole populations, 
and $\Delta \overline{f}_{\mathrm{e,h}}(E,t) = \overline{f}_{\mathrm{e,h}}(E,t) - \overline{f}_{\mathrm{e,h}}(E, -\infty)$ the corresponding excess carrier populations relative to equilibrium. 

\begin{figure}[!t]
\includegraphics[width=\linewidth]{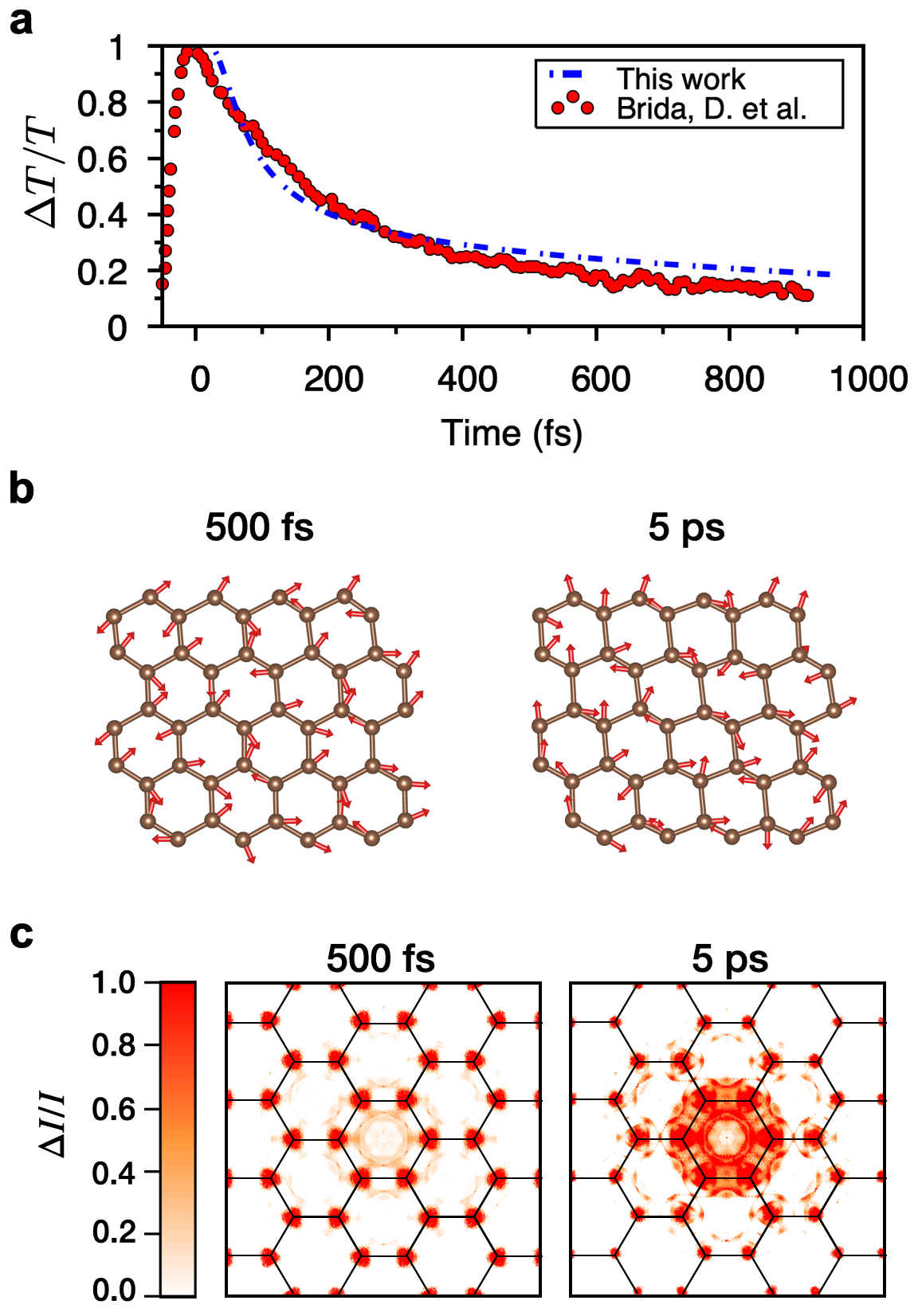}
\caption{\textbf{Simulated ultrafast spectroscopies and structural snapshots in graphene.\\} 
\textbf{a}, Normalized differential transmission $\Delta T/T_0$ at 900 nm wavelength as a function of pump-probe delay time. 
Our calculated result is compared with experiments from Ref.~[\onlinecite{brida2013ultrafast}]. 
\textbf{b}, Structural snapshots at 0.5 ps and 5 ps after pump. 
\textbf{c}, Simulated ultrafast diffuse scattering pattern, $\Delta I(\qq, t)/I$, at 0.5 ps and 5 ps after pump, shown together with the reciprocal lattice.}
\label{figure4}
\end{figure}

Figure~\ref{figure4}a compares the experimental~\cite{brida2013ultrafast} and calculated $\Delta T/T_0$ as a function of pump-probe delay time at 900 nm wavelength. 
The experimental result exhibits two distinct temporal regions. 
At sub-10 fs time delay, $\Delta T/T_0$ increases due to the coupled effects of photoexcitation and electron-electron (e-e) interactions, both of which are not treated explicitly in our calculations but are taken into account in our choice of an initial hot Fermi-Dirac distribution. 
The second time window at $t \!>\! 10$ fs, modeled here explicitly, shows a slow time decay of $\Delta T/T_0$ with a time constant of a few ps. 
Our simulated differential transmission spectrum agrees well with experiment~\cite{brida2013ultrafast} in this time window $-$ following a faster decay of $\Delta T/T_0$ due to carrier cooling in the first 300 fs, we correctly predict the onset of a slower cooling of the excited carriers associated with the bottleneck from slowly rising flexural phonons discussed above. 
While previous work focused on the role of e-e processes~\cite{brida2013ultrafast}, our results show the importance of e-ph interactions at sub-ps times.\\  

\vspace{5pt}
\noindent {\bf Diffuse scattering and structural dynamics.} 
Intense experimental efforts are focused on investigating atomic motions and structural dynamics in the time domain~\cite{lindenberg2017visualization}. 
Among other techniques, ultrafast electron~\cite{Ebi2} and X-ray diffraction~\cite{warren1990x} have made great strides and are currently able to infer atomic displacement patterns and dominant phonon modes governing the nonequilibrium structural dynamics~\cite{lindenberg2017visualization,trigo2010imaging}. 
Using the time-resolved phonon populations for each vibrational mode and wavevector, we can simulate and reconstruct the structural dynamics fully from first principles.
\\
\indent
In a classical picture, the time-dependent displacement of atom $a$ can be decomposed into lattice waves with wavevector $\qq$ and mode $\nu$~\cite{warren1990x},
\begin{equation}
 \mathbf{u}_a(t) = \sum_{\nu \qq} A_{\nu \qq}(t)\,\mathbf{\epsilon}^a_{\nu\qq}e^{-i(\omega_{\nu\qq}t- \qq\cdot \mathbf{r}_a - \delta_{\nu\qq})} \label{eq3}
\end{equation}
where $A_{\nu \qq}$ is the wave amplitude, $\mathbf{\epsilon}^a_{\nu\qq}$ its polarization vector projected on atom $a$, and $\delta_{\nu \qq}$ a phase factor.
For quantized lattice vibrations (phonons), $\mathbf{\epsilon}_{\nu\qq}$ is the phonon eigenvector obtained by diagonalizing the dynamical matrix 
and the amplitude $A_{\nu \qq}$ can be expressed in terms of phonon populations $N_{\nu\qq}$ as
\begin{equation}
A_{\nu \qq}(t) = \sqrt{\frac{\big[2N_{\nu \qq}(t)+1\big]\hbar}{m_a\omega_{\nu\qq}}}\,, 
\label{eq4}
\end{equation}
where $m_a$ is the atomic mass. The phase factors $\delta_{\nu \qq}$ are chosen randomly at time zero and do not affect the dynamics. 
The computed classical atomic displacements $\mathbf{u}_a$ at time delays of 0.5 ps and 5 ps are visualized in Fig.~\ref{figure4}b. 
At 0.5 ps, the excitation of optical phonons from carrier cooling drives a dominant unidirectional vibration of atoms. 
At longer times, as shown in the 5 ps panel, a larger number of phonon modes are incoherently excited through ph-ph processes and the atomic vibrations become randomized, consistent with the thermal phonon distributions we find at 5 ps. 
The quantity measured experimentally in X-ray diffraction is the averaged square atomic displacement, $\braket{\mathbf{u}^2_a} = \sum_{\nu \qq}\frac{1}{2}\braket{A_{\nu \qq}^2}_a$, which we show in Supplementary Video 1, where the radius of the blue sphere around each atom represents $\braket{\mathbf{u}^2_a}$. 
The averaged atomic displacements increase monotonically with time, reaching greater values as the system approaches thermal equilibrium. 
\\
\indent
Thermal diffuse X-ray scattering has proven successful for extracting phonon dispersions and investigating momentum-dependent phonon dynamics~\cite{holt1999determination,stern2018mapping}. 
However, interpreting measurements of the diffuse scattering intensity $I(\mathbf{\qq},t)$ is challenging due to its rich information content. 
For example, mode-resolved phonon contributions cannot be recovered straightforwardly from the measured signal, 
hindering the identification of the dominant phonon modes governing ultrafast structural dynamics and anharmonic phonon processes. 
Here we demonstrate the inverse process of reconstructing the experimental diffuse scattering signal from the computed mode-resolved phonon populations.

\begin{figure}[t]
\includegraphics[width=0.9\linewidth]{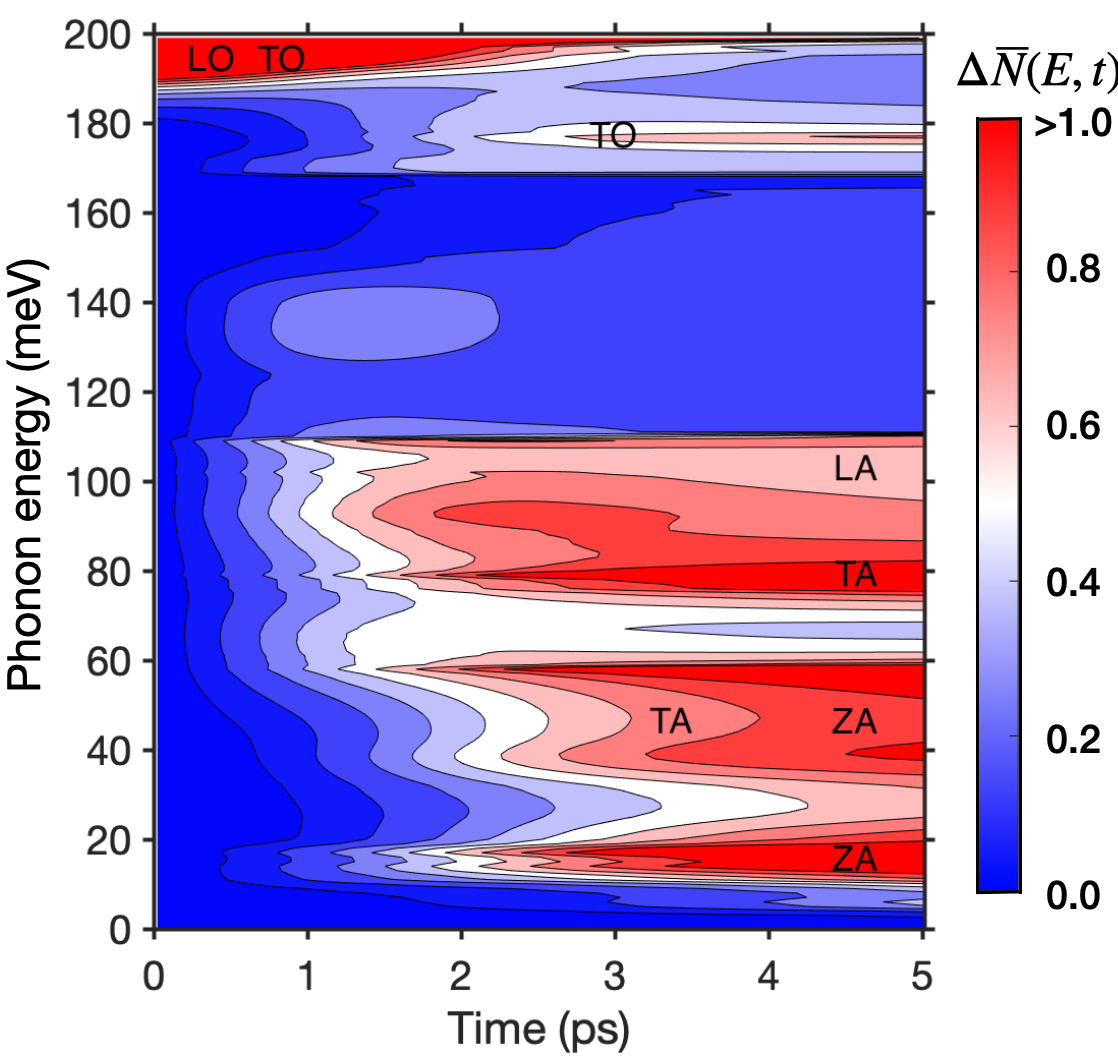}
\caption{Time- and energy-dependent excess phonon populations, $\Delta \overline{N}(E,t)$, after coherent excitation of LO phonons at time zero.} 
\label{figure5}
\end{figure}

We write the scattering intensity at wavevector $\qq$ as~\cite{warren1990x}
\begin{equation}
\label{eq-diffr}
I(\mathbf{\qq},t) \propto e^{-2M}\sum_{\nu}\frac{2\pi^2(2N_{\nu\qq}(t)+1)\hbar(\qq\cdot\mathbf{\epsilon}_{\nu\qq})^2}{m_a\omega_{\nu\qq}}\,,
\end{equation}
where $M$ is the Debye-Waller factor, and compute the ultrafast diffuse scattering signal as the difference between the time-dependent and equilibrium scattering intensities, 
$\Delta I(\qq, t) / I = \frac{I(\qq,t) - I(\qq, -\infty)}{I(\qq, -\infty)}$. 
Our computed diffuse scattering results for graphene are plotted in reciprocal space in Fig.~\ref{figure4}c at 0.5 ps and 5 ps time delays. 
At 0.5 ps, we find peaks of $\Delta I / I$ near the BZ corners due to optical phonons generated from intervalley scattering during the initial sub-ps carrier cooling.  
Small-$\qq$ optical phonons are also generated extensively, as discussed above, but are not visible in Fig.~\ref{figure4}c as the denominator in the differential diffuse scattering $\Delta I / I$ (the thermal equilibrium intensity $I(\qq, -\infty)$) is large at small-$\qq$ due to the large equilibrium population of acoustic phonons. 
The diffuse scattering result at 5 ps captures the dissipation of the optical phonons excess energy to the lower acoustic branches, whereby phonons are generated throughout the BZ.  
While the measured diffuse scattering signal cannot provide information on the mode-resolved dynamics, our simulations can fill this gap, 
providing detailed information on mode-dependent scattering processes in momentum-space and their contributions to diffuse scattering.\\

\noindent{\bf Ultrafast dynamics following phonon excitation.}
Coherent excitation of selected phonon modes has become an important approach for investigating and manipulating materials properties~\cite{rini2007control, graves2013nanoscale, fausti2011light}. 
To demonstrate the flexibility of our numerical framework, we carry out a simulated experiment in which phonons are excited at time zero while electrons and holes are initially kept in their equilibrium room temperature distributions. 
After populating a large excess of LO phonons near the BZ center at time zero, we simulate the subsequent phonon dynamics by solving the coupled electron and phonon BTEs, as above but for different initial conditions. 
The computed time-domain excess phonon populations $\Delta\overline{N}(E,t)$ are plotted in Fig.~\ref{figure5}.  
After 1 ps, LA phonon modes with $\sim$100 meV energy are generated through ph-ph processes, and within 2 ps both LA and TA modes with 80$-$110 meV energy absorb 
most of the excess energy from the optical modes. The flexural ZA phonons come into play only after 3 ps. 
Throughout the simulation, low-energy electron-hole pairs are excited near the Dirac cone via ph-e processes, resulting in a modest ultrasonic attenuation of the phonon dynamics. 
Overall, the excess energy initially imparted to the excited phonons mainly dissipates through slow ph-ph processes, the entire phonon relaxation taking tens of ps. 
Our approach is uniquely able to shed light on these longer timescales characteristic of phonon dynamics. 

\section*{Discussion}
\vspace{-10pt}
A key advantage of our approach is the possibility, rather unique among existing first-principles methods for ultrafast dynamics, to validate the interactions against experiments, thus guaranteeing the quantitative accuracy of our simulated dynamics. We validate the e-ph and ph-ph interactions employed in our calculations, respectively, by computing electrical and heat transport properties of graphene~(see Methods).  
We obtain a room temperature electron mobility of $186,000$ cm$^2$/Vs, in excellent agreement with experimental results in suspended graphene~\cite{G-mobility-1,G-mobility-2}. We also compute the thermal conductivity in the single-mode relaxation time approximation (RTA) of the BTE, obtaining a room temperature value of 482 W/mK in excellent agreement with previous RTA calculations~\cite{fugallo2014thermal}; this result implies that the full solution of the BTE (beyond the RTA) would give a thermal conductivity consistent with experiment~\cite{fugallo2014thermal}. 
These results show that the e-ph and ph-ph interactions employed in our ultrafast dynamics are precise, thus our computed timescales are expected to be quantitatively accurate. 

\section*{Conclusions}
\vspace{-10pt}
In summary, we have shown a versatile numerical framework for modeling the ultrafast coupled dynamics of electrons and phonons. 
We demonstrated its accuracy and broad applicability through simulations of pump-probe spectroscopy, X-ray diffuse scattering, and structural and phonon dynamics. 
Our results shed light on e-ph and ph-ph scattering processes governing ultrafast dynamics in graphene, and provide valuable information for the design of ultrafast electronic and optical devices. 
We plan to make the numerical method available in a future release of {\sc Perturbo}~\cite{zhou2020perturbo} to equip the community with a novel tool for simulating and interpreting ultrafast time-domain experiments. 
Future extensions will aim to treat explicitly the light excitation pulse and the electron spin to unravel the intertwined nonequilibrium dynamics of electronic, structural and spin degrees of freedom. 
Taken together, our first-principles approach demonstrates a paradigm shift in computing the ultrafast electron and atomic vibrational dynamics, bridging the gap between theory and experiment and enabling quantitative predictions of ultrafast phenomena in materials. 
\section*{Methods}
\vspace{-10pt} 

%
\noindent{\bf Density functional theory.}  We carry out first-principles density functional theory (DFT) calculations on graphene with a relaxed in-plane lattice constant of 2.45 $\text{\r{A}}$. 
The graphene sheet is separated from its periodic replicas by a 9 $\text{\r{A}}$ vacuum. The ground-state electronic structure is computed using the {\sc Quantum ESPRESSO} code in the local density approximation (LDA) of DFT. We use a norm-conserving pseudopotential, a 90 Ry plane-wave kinetic energy cutoff, and a Methfessel-Paxton smearing of 0.02 Ry. 
The ground-state charge density is obtained using a 36 $\times$ 36 $\times$ 1 $\kk$-point grid, following which a non-self-consistent calculation is employed to obtain the Kohn-Sham eigenvalues and wave functions on a 12 $\times$ 12 $\times$ 1 $\kk$-point grid. To construct maximally localized Wannier functions (WFs) with the {\sc Wannier90} code~\cite{mostofi2008wannier90}, the Kohn-Sham wave functions are first projected onto atomic $p_z$ orbitals on each atom and $sp^2$ orbitals on every other atom~\cite{marzari2012maximally, jung2013tight}, for a total of 5 wannierized bands.  
The WF spread is then minimized, and the relevant energy windows are adjusted until the interpolated bandstructure can smoothly reproduce the LDA result within $\sim$10 meV throughout the BZ. \\

\vspace{10pt}
\noindent{\bf Electron-phonon scattering and correction to the phonon dispersions.} 
We use density functional perturbation theory~\cite{baroni2001phonons} (DFPT) to compute lattice dynamical properties and the e-ph perturbation potentials, and then form the e-ph matrix elements $g_{nn'\nu}(\kk,\qq)$ on coarse 12 $\times$ 12 $\times$ 1 $\kk$-point and $\qq$-point BZ grids using {\sc Perturbo}~\cite{zhou2020perturbo}. 
Here and below, the e-ph matrix elements $g_{nn'\nu}(\kk,\qq)$ quantify the probability amplitude for an electron in Bloch state $\ket{n\kk}$ with energy $E_{n\kk}$ to scatter into a final state $\ket{n'\kk+\qq}$ with energy $E_{n'\kk+\qq}$ due to emission or absorption of a phonon with branch index $\nu$, wavevector $\qq$, and energy $\hbar\omega_{\nu\qq}$. 
The electron and phonon energies and the e-ph matrix elements are interpolated on fine grids using WFs with {\sc Perturbo}~\cite{zhou2020perturbo}. The average e-ph coupling strengths $\overline{ g^2_{\nu}}$ used in Fig.~\ref{figure3}a are obtained as $\overline{ g^2_{\nu}} = \sum_{nn'\kk\qq}|g_{nn'\nu}(\kk,\qq)|^2$, where the summation includes electronic states near the Dirac point and phonon wavevectors in the entire BZ.
\\
\indent
To account for the Kohn anomaly near $\qq = $ K~\cite{piscanec2004kohn}, we interpolate $E_{n\kk}$ and $g_{nn'\nu}(\kk,\qq)$ on a random BZ grid of $10^4$ points, 
and obtain graphene phonon dispersions $\omega_{\nu \qq}$ that include a previously proposed GW correction~\cite{lazzeri2008impact},

\begin{equation}
\omega_{\nu \qq} = \sqrt{\frac{B_\qq^{GW}}{m} + \frac{4\gamma^{GW}_{\qq}}{\mathcal{N}'_\kk}\sum_{\kk}\frac{|g_{nn'\nu}(\kk,\qq)|^2}{E_{\pi\kk} - E_{\pi^*\kk+\qq}}}
\end{equation}
where $\pi$($\pi^*$) labels the occupied (emtpy) $\pi$ band, $\mathcal{N}'_\kk = 10^4$ is the total number of $\kk$-points in the random grid, 
and $B_\qq^{GW}$ is a parameter employed to converge the LO and TO phonon energies at K. The rescaling factor $\gamma^{GW}_{\qq}$ is computed using~\cite{venezuela2011theory}
\begin{equation}
\gamma^{GW}_{\qq} = 1 + (\gamma^{GW} - 1)\frac{1}{2}\text{erfc}\left(\frac{|\qq-\text{K}^n|\frac{a_0}{2\pi}-0.2}{0.05}\right)
\end{equation}
with $a_0$ the graphene lattice constant, $\gamma^{GW}$ a constant equal to 1.61, and K$^n$ the nearest vector to $\qq$ among those equivalent to K.
\\
\indent
To verify the convergence of the $\qq$-point grid and compute the electrical mobility, we use {\sc Perturbo}~\cite{zhou2020perturbo} to calculate the state-dependent e-ph scattering rates $\Gamma_{n\kk}$ within lowest-order perturbation theory~\cite{bernardi2016first},
\begin{align}
\Gamma_{n\kk} = &\frac{2\pi}{\hbar} \frac{1}{\mathcal{N}_\qq} \sum_{n'\qq\nu}|g_{nn'\nu}(\kk,\qq)|^2& \nonumber\\
&\times [(N_{\nu\qq} + 1 - f_{n'\kk+\qq})\delta(E_{n\kk}-\hbar\omega_{\nu \qq} - E_{n'\kk+\qq})& \nonumber\\
&+ (N_{\nu\qq} + f_{n'\kk+\qq})\delta(E_{n\kk} +\hbar \omega_{\nu \qq} - E_{n'\kk+\qq})]\,,&
\label{e-lifetime}
\end{align}
where $f_{n\kk}$ and $N_{\nu\qq}$ are the electron and phonon equilibrium occupations at 300 K. Convergence is achieved for a grid of 420 $\times$ 420 $\times$ 1 $\qq$-points (see Supplementary Figure S1) using a Gaussian broadening of 20 meV to approximate the $\delta$ functions in Eq.~(\ref{e-lifetime}). The same grid is employed for all ultrafast dynamics calculations.\\

\vspace{8pt}
\noindent{\bf Phonon-phonon scattering.} We use the temperature-dependent effective
potential (TDEP) method~\cite{hellman2011lattice, hellman2013temperature, hellman2013temperature2} to obtain the interatomic force constants and ph-ph matrix elements $\Phi_{\nu\nu'\nu''}(\qq,\qq')$. The latter are the probability amplitudes for three-phonon scattering processes in which the phonon state $\ket{\nu\qq}$ with energy $\hbar\omega_{\nu\qq}$ 
scatters to the states $\ket{\nu''\qq+\qq'}$ and $\ket{\nu' \qq'}$, or the inverse process in which the last two phonons combine to generate $\ket{\nu \qq}$. 
For the TDEP calculations, we prepare a number of 12 $\times$ 12 $\times$ 1 (288 atom) supercells with random thermal displacements corresponding to a canonical ensemble at 300 K. 
To validate the ph-ph interactions and compute the thermal conductivity, we compute the phonon scattering rates $\Lambda_{\nu\qq}$ as~\cite{debernardi1995anharmonic} 
\begin{align}
\Lambda_{\nu\qq} = &\frac{18\pi}{\hbar^2}\,\, \frac{1}{\mathcal{N}^2_\qq} \sum_{\nu'\nu'' \qq' \qq''} |\Phi_{\nu\nu'\nu''}(\qq',\qq'')|^2& \nonumber\\
&\times [(N_{\nu'\qq'} + N_{\nu''\qq''} + 1)\delta(\omega_{\nu \qq}-\omega_{\nu' \qq'} - \omega_{\nu \qq''})& \nonumber\\
&+ 2(N_{\nu'\qq'} - N_{\nu''\qq''})\delta(\omega_{\nu \qq}-\omega_{\nu' \qq'} + \omega_{\nu \qq''})]&
\label{ph-lifetime}
\end{align}
using phonon equilibrium occupations at 300 K. We compute and converge the ph-ph scattering rates (see Supplementary Figure S2) with an in-house version of {\sc Perturbo}~\cite{zhou2020perturbo}. The ph-ph scattering rates converge for grids equal to or finer than 210 $\times$ 210 $\times$ 1 $\qq$-points using a 1 meV Gaussian broadening to approximate the $\delta$ functions in Eq.~(\ref{ph-lifetime}).\\

\noindent{\bf Coupled electron and phonon ultrafast dynamics.} 
We simulate the ultrafast dynamics of coupled electrons and phonons due to e-ph and ph-ph scattering processes in graphene in the absence of external fields. 
The time evolution of the carrier and phonon distributions is obtained by solving the coupled carrier and phonon BTEs:

\begin{widetext}
      \begin{align}
        \frac{\partial f_{n\kk}(t)}{\partial t} = & -\frac{2\pi}{\hbar}\frac{1}{\mathcal{N}_\qq} \sum_{n' \qq \nu}|g_{nn'\nu}(\kk,\qq)|^2 \{\delta(E_{n\kk} - E_{n'\kk+\qq} -\hbar \omega_{\nu \qq}) [f_{n\kk}(1-f_{n'\kk+\qq})(N_{\nu \qq} + 1) - f_{n'\kk+\qq}(1-f_{n\kk})N_{\nu\qq}]& \nonumber\\
        &+\delta(E_{n\kk} - E_{n'\kk+\qq} + \hbar \omega_{\nu \qq}) [f_{n\kk}(1-f_{n'\kk+\qq})N_{\nu \qq} - f_{n'\kk+\qq}(1-f_{n\kk})(N_{\nu\qq}+1)]\}& \nonumber\\
        \frac{\partial N_{\nu\qq}(t)}{\partial t} =& -\frac{4\pi}{\hbar}\frac{1}{\mathcal{N}_\kk} \sum_{n' \kk \nu}|g_{nn'\nu}(\kk,\qq)|^2 \delta(E_{n\kk} - E_{n'\kk+\qq} 
        + \hbar \omega_{\nu \qq}
        [f_{n\kk}(1-f_{n'\kk+\qq})N_{\nu \qq} - f_{n'\kk+\qq}(1-f_{n\kk})(N_{\nu\qq}+1)]& \nonumber \\
         &-\frac{18\pi}{\hbar^2}\frac{1}{\mathcal{N}_\qq^2} \sum_{\nu'\nu'' \qq' \qq''}|\Phi_{\nu\nu'\nu''}(\qq',\qq'')|^2\{\delta(\omega_{\nu\qq} - \omega_{\nu'\qq'} - \omega_{\nu'' \qq''}) [N_{\nu\qq}(N_{\nu'\qq'}+1)(N_{\nu'' \qq''}+1) - N_{\nu'\qq'}N_{\nu''\qq''}(N_{\nu\qq}+1)]& \nonumber\\
&+\delta(\omega_{\nu\qq} + \omega_{\nu'\qq'} - \omega_{\nu'' \qq''})[N_{\nu\qq}N_{\nu'\qq'}(N_{\nu'' \qq''}+1) - N_{\nu'\qq'}(N_{\nu''\qq''}+1)(N_{\nu\qq}+1)]\}& \label{BTEs}\,,
      \end{align}
\end{widetext}
which account for e-ph scattering for electrons, e-ph scattering for phonons (denoted above as ph-e scattering) and ph-ph scattering for phonons. 
In Eq.~(\ref{BTEs}), $f_{n\kk}(t)$ are time-dependent electron populations and $N_{\nu \qq}(t)$ are time-dependent phonon populations. The e-ph matrix elements $g_{nn'\nu}(\kk,\qq)$ and ph-ph matrix elements $\Phi_{\nu\nu'\nu''}(\qq',\qq'')$ are computed using DFPT in the ground state plus WF interpolation and TDEP, respectively, as specified above. 
\\
\indent
We numerically solve Eq. (\ref{BTEs}) using the fourth-order Runge-Kutta method with a time step of 2 fs, 
using uniform fine BZ grids of 420 $\times$ 420 $\times$ 1 $\kk$- and $\qq$-points for all e-ph and ph-ph scattering processes. 
The ph-ph scattering matrix elements are computed on a uniform 210 $\times$ 210 $\times$ 1 $\qq$-point BZ grid and then Fourier interpolated to a 420 $\times$ 420 $\times$ 1 $\qq$-point grid. 
These grids are sufficient to converge the scattering processes and ultrafast dynamics. To reduce the number of scattering processes entering the BTEs, we use only electronic states in the energy window of relevance for our calculations, which restricts the number of $\kk$-points to a few thousands, while the $\qq$-points span the entire BZ. 
We additionally select the relevant ph-ph processes by imposing energy conservation (within a few times the broadening value) thereby dramatically reducing the number of ph-ph scattering processes, which would otherwise be unmanageable for the fine $\qq$-point grid employed. 
The BZ-averaged energy-dependent carrier populations $\overline{f}(E,t)$ and phonon populations $\overline{N}(E,t)$ at energy $E$, first used in Fig.~\ref{RTS}, are obtained respectively 
as $\overline{f}(E, t) = \sum_{n\kk}f_{n\kk}(t) \delta(E_{n\kk} – E)$ and $\overline{N}(E, t) = \sum_{\nu\qq}N_{\nu\qq}(t) \delta(\hbar\omega_{\nu\qq} – E)$ via tetrahedron integration. 
The average temperature of each phonon mode at time $t$ (see Fig.~\ref{figure3}) is computed as $\overline{T}_\nu (t) = \sum_\qq T_{\nu\qq}(t)$, where the state-dependent temperature $T_{\nu\qq}(t)$ is obtained by inverting the Bose-Einstein occupation formula, $N_{\nu \qq}(t) = (e^{\hbar \omega_{\nu \qq} / k_BT_{\nu \qq}(t)} - 1)^{-1}$, using the phonon populations at each time step. 
The phonon populations follow a Bose-Einstein thermal distribution at the computed temperature $\overline{T}_\nu (t)$ for times greater than 2 ps for all acoustic modes and the ZO mode,  
and at times greater than 4$-$5 ps for the LO and TO optical modes. 
\\
\indent
We develop a scheme combining MPI and {\sc openMP} parallelization to efficiently compute the e-ph, ph-e and ph-ph collision integrals at each time step. 
Briefly, $(\kk,\qq)$ pairs for e-ph scattering and $(\qq,\qq',\qq'')$ triplets for ph-ph scattering are distributed among MPI processes; 
using {\sc openMP} parallelization, rapid on-node operations are employed to compute the collision integrals within each MPI process; the results are then added together to form the collision integrals.   
This scheme leverages an algorithm we previously employed to time-step the electron BTE~\cite{jhalani2017ultrafast,zhou2020perturbo}; however, including ph-ph scattering is dramatically more difficult due to the need to take into account all $\qq$-point triplets involved in the ph-ph processes. To accomplish this task, we developed a new parallel implementation of the phonon BTE in this work.\\

\noindent{\bf Transport properties.} The electron mobility is computed from a full solution of the linearized electron BTE with the {\sc Perturbo} code~\cite{zhou2020perturbo}. The thermal conductivity is computed using ph-ph matrix elements from TDEP with an in-house implementation of the single-mode RTA formula~\cite{fugallo2014thermal}. 
Both calculations use the converged BZ grids given above.
\vspace{-16pt}
\section*{Data availability}
\vspace{-12pt} 
The datasets generated and/or analyzed in the current study are available from the corresponding author upon reasonable request.

\vspace{20pt}
{\bf \large References}
\vspace{-20pt}
\bibliographystyle{nature}

\begin{thebibliography}{10}
\expandafter\ifx\csname url\endcsname\relax
  \def\url#1{\texttt{#1}}\fi
\expandafter\ifx\csname urlprefix\endcsname\relax\def\urlprefix{URL }\fi
\providecommand{\bibinfo}[2]{#2}
\providecommand{\eprint}[2][]{\url{#2}}

\bibitem{buzzi2018probing}
\bibinfo{author}{Buzzi, M.}, \bibinfo{author}{F{\"o}rst, M.},
  \bibinfo{author}{Mankowsky, R.} \& \bibinfo{author}{Cavalleri, A.}
\newblock \bibinfo{title}{Probing dynamics in quantum materials with
  femtosecond {X}-rays}.
\newblock \emph{\bibinfo{journal}{Nat. Rev. Mater.}}
  \textbf{\bibinfo{volume}{3}}, \bibinfo{pages}{299} (\bibinfo{year}{2018}).
\newblock \urlprefix\url{https://www.nature.com/articles/s41578-018-0024-9}.

\bibitem{najafi2017super}
\bibinfo{author}{Najafi, E.}, \bibinfo{author}{Ivanov, V.},
  \bibinfo{author}{Zewail, A.} \& \bibinfo{author}{Bernardi, M.}
\newblock \bibinfo{title}{Super-diffusion of excited carriers in
  semiconductors}.
\newblock \emph{\bibinfo{journal}{Nat. Commun.}} \textbf{\bibinfo{volume}{8}},
  \bibinfo{pages}{15177} (\bibinfo{year}{2017}).
\newblock \urlprefix\url{https://www.nature.com/articles/ncomms15177}.

\bibitem{zurch2017ultrafast}
\bibinfo{author}{Z{\"u}rch, M.} \emph{et~al.}
\newblock \bibinfo{title}{Ultrafast carrier thermalization and trapping in
  silicon-germanium alloy probed by extreme ultraviolet transient absorption
  spectroscopy}.
\newblock \emph{\bibinfo{journal}{Struct. Dyn.}} \textbf{\bibinfo{volume}{4}},
  \bibinfo{pages}{044029} (\bibinfo{year}{2017}).
\newblock \urlprefix\url{https://aca.scitation.org/doi/abs/10.1063/1.4985056}.

\bibitem{Damascelli2019}
\bibinfo{author}{Na, M.~X.} \emph{et~al.}
\newblock \bibinfo{title}{Direct determination of mode-projected
  electron-phonon coupling in the time domain}.
\newblock \emph{\bibinfo{journal}{Science}} \textbf{\bibinfo{volume}{366}},
  \bibinfo{pages}{1231--1236} (\bibinfo{year}{2019}).
\newblock \urlprefix\url{https://science.sciencemag.org/content/366/6470/1231}.

\bibitem{young2010femtosecond}
\bibinfo{author}{Young, L.} \emph{et~al.}
\newblock \bibinfo{title}{Femtosecond electronic response of atoms to
  ultra-intense {X}-rays}.
\newblock \emph{\bibinfo{journal}{Nature}} \textbf{\bibinfo{volume}{466}},
  \bibinfo{pages}{56} (\bibinfo{year}{2010}).
\newblock \urlprefix\url{https://www.nature.com/articles/nature09177}.

\bibitem{fritz2007ultrafast}
\bibinfo{author}{Fritz, D.~M.} \emph{et~al.}
\newblock \bibinfo{title}{Ultrafast bond softening in bismuth: mapping a
  solid's interatomic potential with {X}-rays}.
\newblock \emph{\bibinfo{journal}{Science}} \textbf{\bibinfo{volume}{315}},
  \bibinfo{pages}{633--636} (\bibinfo{year}{2007}).
\newblock \urlprefix\url{https://science.sciencemag.org/content/315/5812/633}.

\bibitem{reis2006ultrafast}
\bibinfo{author}{Reis, D.~A.} \& \bibinfo{author}{Lindenberg, A.~M.}
\newblock \bibinfo{title}{Ultrafast {X}-ray scattering in solids}.
\newblock In \emph{\bibinfo{booktitle}{{Light Scattering in Solid IX}}},
  \bibinfo{pages}{371--422} (\bibinfo{publisher}{Springer},
  \bibinfo{year}{2006}).

\bibitem{trigo2013fourier}
\bibinfo{author}{Trigo, M.} \emph{et~al.}
\newblock \bibinfo{title}{Fourier-transform inelastic {X}-ray scattering from
  time-and momentum-dependent phonon--phonon correlations}.
\newblock \emph{\bibinfo{journal}{Nat. Phys.}} \textbf{\bibinfo{volume}{9}},
  \bibinfo{pages}{790} (\bibinfo{year}{2013}).
\newblock \urlprefix\url{https://www.nature.com/articles/nphys2788}.

\bibitem{lindenberg2000time}
\bibinfo{author}{Lindenberg, A.} \emph{et~al.}
\newblock \bibinfo{title}{Time-resolved {X}-ray diffraction from coherent
  phonons during a laser-induced phase transition}.
\newblock \emph{\bibinfo{journal}{Phys. Rev. Lett.}}
  \textbf{\bibinfo{volume}{84}}, \bibinfo{pages}{111} (\bibinfo{year}{2000}).
\newblock
  \urlprefix\url{https://journals.aps.org/prl/abstract/10.1103/PhysRevLett.84.111}.

\bibitem{zalden2019femtosecond}
\bibinfo{author}{Zalden, P.} \emph{et~al.}
\newblock \bibinfo{title}{Femtosecond {X}-ray diffraction reveals a
  liquid--liquid phase transition in phase-change materials}.
\newblock \emph{\bibinfo{journal}{Science}} \textbf{\bibinfo{volume}{364}},
  \bibinfo{pages}{1062--1067} (\bibinfo{year}{2019}).
\newblock \urlprefix\url{https://science.sciencemag.org/content/364/6445/1062}.

\bibitem{de2013speed}
\bibinfo{author}{De~Jong, S.} \emph{et~al.}
\newblock \bibinfo{title}{Speed limit of the insulator--metal transition in
  magnetite}.
\newblock \emph{\bibinfo{journal}{Nat. Mater.}} \textbf{\bibinfo{volume}{12}},
  \bibinfo{pages}{882} (\bibinfo{year}{2013}).
\newblock \urlprefix\url{https://www.nature.com/articles/nmat3718}.

\bibitem{mankowsky2017ultrafast}
\bibinfo{author}{Mankowsky, R.}, \bibinfo{author}{von Hoegen, A.},
  \bibinfo{author}{F{\"o}rst, M.} \& \bibinfo{author}{Cavalleri, A.}
\newblock \bibinfo{title}{Ultrafast reversal of the ferroelectric
  polarization}.
\newblock \emph{\bibinfo{journal}{Phys. Rev. Lett.}}
  \textbf{\bibinfo{volume}{118}}, \bibinfo{pages}{197601}
  (\bibinfo{year}{2017}).
\newblock
  \urlprefix\url{https://journals.aps.org/prl/abstract/10.1103/PhysRevLett.118.197601}.

\bibitem{mitrano2016possible}
\bibinfo{author}{Mitrano, M.} \emph{et~al.}
\newblock \bibinfo{title}{Possible light-induced superconductivity in
  {K$_3$C$_{60}$} at high temperature}.
\newblock \emph{\bibinfo{journal}{Nature}} \textbf{\bibinfo{volume}{530}},
  \bibinfo{pages}{461} (\bibinfo{year}{2016}).
\newblock \urlprefix\url{https://www.nature.com/articles/nature16522}.

\bibitem{wilson2013two}
\bibinfo{author}{Wilson, R.}, \bibinfo{author}{Feser, J.~P.},
  \bibinfo{author}{Hohensee, G.~T.} \& \bibinfo{author}{Cahill, D.~G.}
\newblock \bibinfo{title}{Two-channel model for nonequilibrium thermal
  transport in pump-probe experiments}.
\newblock \emph{\bibinfo{journal}{Phys. Rev. B}} \textbf{\bibinfo{volume}{88}},
  \bibinfo{pages}{144305} (\bibinfo{year}{2013}).
\newblock
  \urlprefix\url{https://journals.aps.org/prb/abstract/10.1103/PhysRevB.88.144305}.

\bibitem{haug2008quantum}
\bibinfo{author}{Haug, H.} \& \bibinfo{author}{Jauho, A.-P.}
\newblock \emph{\bibinfo{title}{Quantum kinetics in transport and optics of
  semiconductors}}, vol.~\bibinfo{volume}{2} (\bibinfo{publisher}{Springer
  Science \& Business Media, Boston}, \bibinfo{year}{2008}).

\bibitem{waldecker2016electron}
\bibinfo{author}{Waldecker, L.}, \bibinfo{author}{Bertoni, R.},
  \bibinfo{author}{Ernstorfer, R.} \& \bibinfo{author}{Vorberger, J.}
\newblock \bibinfo{title}{Electron-phonon coupling and energy flow in a simple
  metal beyond the two-temperature approximation}.
\newblock \emph{\bibinfo{journal}{Phys. Rev. X}} \textbf{\bibinfo{volume}{6}},
  \bibinfo{pages}{021003} (\bibinfo{year}{2016}).
\newblock
  \urlprefix\url{https://journals.aps.org/prx/abstract/10.1103/PhysRevX.6.021003}.

\bibitem{murray2007phonon}
\bibinfo{author}{Murray, E.} \emph{et~al.}
\newblock \bibinfo{title}{Phonon dispersion relations and softening in
  photoexcited bismuth from first principles}.
\newblock \emph{\bibinfo{journal}{Phys. Rev. B}} \textbf{\bibinfo{volume}{75}},
  \bibinfo{pages}{184301} (\bibinfo{year}{2007}).
\newblock
  \urlprefix\url{https://journals.aps.org/prb/abstract/10.1103/PhysRevB.75.184301}.

\bibitem{Bernardi-Si}
\bibinfo{author}{Bernardi, M.}, \bibinfo{author}{Vigil-Fowler, D.},
  \bibinfo{author}{Lischner, J.}, \bibinfo{author}{Neaton, J.~B.} \&
  \bibinfo{author}{Louie, S.~G.}
\newblock \bibinfo{title}{Ab initio study of hot carriers in the first
  picosecond after sunlight absorption in silicon}.
\newblock \emph{\bibinfo{journal}{Phys. Rev. Lett.}}
  \textbf{\bibinfo{volume}{112}}, \bibinfo{pages}{257402}
  (\bibinfo{year}{2014}).
\newblock
  \urlprefix\url{http://link.aps.org/doi/10.1103/PhysRevLett.112.257402}.

\bibitem{Bernardi-SPP}
\bibinfo{author}{Bernardi, M.}, \bibinfo{author}{Mustafa, J.},
  \bibinfo{author}{Neaton, J.~B.} \& \bibinfo{author}{Louie, S.~G.}
\newblock \bibinfo{title}{Theory and computation of hot carriers generated by
  surface plasmon polaritons in noble metals}.
\newblock \emph{\bibinfo{journal}{Nat. Commun.}} \textbf{\bibinfo{volume}{6}}
  (\bibinfo{year}{2015}).
\newblock \urlprefix\url{http://dx.doi.org/10.1038/ncomms8044}.

\bibitem{jhalani2017ultrafast}
\bibinfo{author}{Jhalani, V.~A.}, \bibinfo{author}{Zhou, J.-J.} \&
  \bibinfo{author}{Bernardi, M.}
\newblock \bibinfo{title}{Ultrafast hot carrier dynamics in {GaN} and its
  impact on the efficiency droop}.
\newblock \emph{\bibinfo{journal}{Nano Lett.}} \textbf{\bibinfo{volume}{17}},
  \bibinfo{pages}{5012--5019} (\bibinfo{year}{2017}).
\newblock
  \urlprefix\url{https://pubs.acs.org/doi/abs/10.1021/acs.nanolett.7b02212}.

\bibitem{bernardi2016first}
\bibinfo{author}{Bernardi, M.}
\newblock \bibinfo{title}{First-principles dynamics of electrons and phonons}.
\newblock \emph{\bibinfo{journal}{Eur. Phys. J. B}}
  \textbf{\bibinfo{volume}{89}}, \bibinfo{pages}{239} (\bibinfo{year}{2016}).
\newblock \urlprefix\url{https://doi.org/10.1140/epjb/e2016-70399-4}.

\bibitem{sadasivam2017theory}
\bibinfo{author}{Sadasivam, S.}, \bibinfo{author}{Chan, M.~K.} \&
  \bibinfo{author}{Darancet, P.}
\newblock \bibinfo{title}{Theory of thermal relaxation of electrons in
  semiconductors}.
\newblock \emph{\bibinfo{journal}{Phys. Rev. Lett.}}
  \textbf{\bibinfo{volume}{119}}, \bibinfo{pages}{136602}
  (\bibinfo{year}{2017}).
\newblock
  \urlprefix\url{https://journals.aps.org/prl/abstract/10.1103/PhysRevLett.119.136602}.

\bibitem{yabana1996time}
\bibinfo{author}{Yabana, K.} \& \bibinfo{author}{Bertsch, G.}
\newblock \bibinfo{title}{Time-dependent local-density approximation in real
  time}.
\newblock \emph{\bibinfo{journal}{Phys. Rev. B}} \textbf{\bibinfo{volume}{54}},
  \bibinfo{pages}{4484} (\bibinfo{year}{1996}).
\newblock
  \urlprefix\url{https://journals.aps.org/prb/abstract/10.1103/PhysRevB.54.4484}.

\bibitem{castro2004propagators}
\bibinfo{author}{Castro, A.}, \bibinfo{author}{Marques, M.~A.} \&
  \bibinfo{author}{Rubio, A.}
\newblock \bibinfo{title}{Propagators for the time-dependent {Kohn--Sham}
  equations}.
\newblock \emph{\bibinfo{journal}{J. Chem. Phys.}}
  \textbf{\bibinfo{volume}{121}}, \bibinfo{pages}{3425--3433}
  (\bibinfo{year}{2004}).
\newblock \urlprefix\url{https://aip.scitation.org/doi/10.1063/1.1774980}.

\bibitem{marques2004time}
\bibinfo{author}{Marques, M.~A.} \& \bibinfo{author}{Gross, E.~K.}
\newblock \bibinfo{title}{Time-dependent density functional theory}.
\newblock \emph{\bibinfo{journal}{Annu. Rev. Phys. Chem.}}
  \textbf{\bibinfo{volume}{55}}, \bibinfo{pages}{427--455}
  (\bibinfo{year}{2004}).
\newblock
  \urlprefix\url{https://www.annualreviews.org/doi/abs/10.1146/annurev.physchem.55.091602.094449}.

\bibitem{nazeeruddin2005combined}
\bibinfo{author}{Nazeeruddin, M.~K.} \emph{et~al.}
\newblock \bibinfo{title}{Combined experimental and {DFT-TDDFT} computational
  study of photoelectrochemical cell ruthenium sensitizers}.
\newblock \emph{\bibinfo{journal}{J. Am. Chem. Soc.}}
  \textbf{\bibinfo{volume}{127}}, \bibinfo{pages}{16835--16847}
  (\bibinfo{year}{2005}).
\newblock \urlprefix\url{https://pubs.acs.org/doi/abs/10.1021/ja052467l}.

\bibitem{rozzi2013quantum}
\bibinfo{author}{Rozzi, C.~A.} \emph{et~al.}
\newblock \bibinfo{title}{Quantum coherence controls the charge separation in a
  prototypical artificial light-harvesting system}.
\newblock \emph{\bibinfo{journal}{Nat. Commun.}} \textbf{\bibinfo{volume}{4}},
  \bibinfo{pages}{1602} (\bibinfo{year}{2013}).
\newblock \urlprefix\url{https://www.nature.com/articles/ncomms2603}.

\bibitem{sottile2005tddft}
\bibinfo{author}{Sottile, F.} \emph{et~al.}
\newblock \bibinfo{title}{{TDDFT }from molecules to solids: The role of
  long-range interactions}.
\newblock \emph{\bibinfo{journal}{Int. J. Quantum Chem.}}
  \textbf{\bibinfo{volume}{102}}, \bibinfo{pages}{684--701}
  (\bibinfo{year}{2005}).
\newblock
  \urlprefix\url{https://onlinelibrary.wiley.com/doi/full/10.1002/qua.20486}.

\bibitem{marques2012fundamentals}
\bibinfo{author}{Marques, M.~A.}, \bibinfo{author}{Maitra, N.~T.},
  \bibinfo{author}{Nogueira, F.~M.}, \bibinfo{author}{Gross, E.~K.} \&
  \bibinfo{author}{Rubio, A.}
\newblock \emph{\bibinfo{title}{Fundamentals of Time-dependent Density
  Functional Theory}}, vol. \bibinfo{volume}{837} (\bibinfo{publisher}{Springer
  Science \& Business Media, Boston}, \bibinfo{year}{2012}).

\bibitem{HeadGordon-2003}
\bibinfo{author}{Dreuw, A.}, \bibinfo{author}{Weisman, J.~L.} \&
  \bibinfo{author}{Head-Gordon, M.}
\newblock \bibinfo{title}{Long-range charge-transfer excited states in
  time-dependent density functional theory require non-local exchange}.
\newblock \emph{\bibinfo{journal}{J. Chem. Phys.}}
  \textbf{\bibinfo{volume}{119}}, \bibinfo{pages}{2943--2946}
  (\bibinfo{year}{2003}).
\newblock \urlprefix\url{https://aip.scitation.org/doi/10.1063/1.1590951}.

\bibitem{Maitra-2012}
\bibinfo{author}{Elliott, P.}, \bibinfo{author}{Fuks, J.~I.},
  \bibinfo{author}{Rubio, A.} \& \bibinfo{author}{Maitra, N.~T.}
\newblock \bibinfo{title}{Universal dynamical steps in the exact time-dependent
  exchange-correlation potential}.
\newblock \emph{\bibinfo{journal}{Phys. Rev. Lett.}}
  \textbf{\bibinfo{volume}{109}}, \bibinfo{pages}{266404}
  (\bibinfo{year}{2012}).
\newblock
  \urlprefix\url{https://link.aps.org/doi/10.1103/PhysRevLett.109.266404}.

\bibitem{yang2020simultaneous}
\bibinfo{author}{Yang, J.} \emph{et~al.}
\newblock \bibinfo{title}{Simultaneous observation of nuclear and electronic
  dynamics by ultrafast electron diffraction}.
\newblock \emph{\bibinfo{journal}{Science}} \textbf{\bibinfo{volume}{368}},
  \bibinfo{pages}{885--889} (\bibinfo{year}{2020}).
\newblock \urlprefix\url{https://science.sciencemag.org/content/368/6493/885}.

\bibitem{Mahan-main}
\bibinfo{author}{Mahan, G.~D.}
\newblock \emph{\bibinfo{title}{Many-Particle Physics}}
  (\bibinfo{publisher}{Springer Science \& Business Media, Boston},
  \bibinfo{year}{2000}).

\bibitem{sangalli2015complete}
\bibinfo{author}{{Sangalli, D.}} \& \bibinfo{author}{{Marini, A.}}
\newblock \bibinfo{title}{Ultra-fast carriers relaxation in bulk silicon
  following photo-excitation with a short and polarized laser pulse}.
\newblock \emph{\bibinfo{journal}{Eur. Phys. Lett.}}
  \textbf{\bibinfo{volume}{110}}, \bibinfo{pages}{47004}
  (\bibinfo{year}{2015}).
\newblock \urlprefix\url{http://dx.doi.org/10.1209/0295-5075/110/47004}.

\bibitem{Mahan-nutshell}
\bibinfo{author}{Mahan, G.~D.}
\newblock \emph{\bibinfo{title}{Condensed Matter in a Nutshell}}
  (\bibinfo{publisher}{Princeton University Press, Princeton},
  \bibinfo{year}{2010}).

\bibitem{zhou2020perturbo}
\bibinfo{author}{Zhou, J.-J.} \emph{et~al.}
\newblock \bibinfo{title}{Perturbo: a software package for ab initio
  electron-phonon interactions, charge transport and ultrafast dynamics}.
\newblock \emph{\bibinfo{journal}{arXiv: 2002.02045}}  (\bibinfo{year}{2020}).
\newblock \urlprefix\url{https://arxiv.org/abs/2002.02045}.

\bibitem{dawlaty2008measurement}
\bibinfo{author}{Dawlaty, J.~M.}, \bibinfo{author}{Shivaraman, S.},
  \bibinfo{author}{Chandrashekhar, M.}, \bibinfo{author}{Rana, F.} \&
  \bibinfo{author}{Spencer, M.~G.}
\newblock \bibinfo{title}{Measurement of ultrafast carrier dynamics in
  epitaxial graphene}.
\newblock \emph{\bibinfo{journal}{Appl. Phys. Lett.}}
  \textbf{\bibinfo{volume}{92}}, \bibinfo{pages}{042116}
  (\bibinfo{year}{2008}).
\newblock \urlprefix\url{https://aip.scitation.org/doi/abs/10.1063/1.2837539}.

\bibitem{winzer2012impact}
\bibinfo{author}{Winzer, T.} \& \bibinfo{author}{Mali{\'c}, E.}
\newblock \bibinfo{title}{Impact of auger processes on carrier dynamics in
  graphene}.
\newblock \emph{\bibinfo{journal}{Phys. Rev. B}} \textbf{\bibinfo{volume}{85}},
  \bibinfo{pages}{241404} (\bibinfo{year}{2012}).
\newblock
  \urlprefix\url{https://journals.aps.org/prb/abstract/10.1103/PhysRevB.85.241404}.

\bibitem{wang2010ultrafast}
\bibinfo{author}{Wang, H.} \emph{et~al.}
\newblock \bibinfo{title}{Ultrafast relaxation dynamics of hot optical phonons
  in graphene}.
\newblock \emph{\bibinfo{journal}{Appl. Phys. Lett.}}
  \textbf{\bibinfo{volume}{96}}, \bibinfo{pages}{081917}
  (\bibinfo{year}{2010}).
\newblock
  \urlprefix\url{https://aip.scitation.org/doi/figure/10.1063/1.3291615}.

\bibitem{sun2012dynamics}
\bibinfo{author}{Sun, B.}, \bibinfo{author}{Zhou, Y.} \& \bibinfo{author}{Wu,
  M.}
\newblock \bibinfo{title}{Dynamics of photoexcited carriers in graphene}.
\newblock \emph{\bibinfo{journal}{Phys. Rev. B}} \textbf{\bibinfo{volume}{85}},
  \bibinfo{pages}{125413} (\bibinfo{year}{2012}).
\newblock
  \urlprefix\url{https://journals.aps.org/prb/abstract/10.1103/PhysRevB.85.125413}.

\bibitem{gierz2013snapshots}
\bibinfo{author}{Gierz, I.} \emph{et~al.}
\newblock \bibinfo{title}{Snapshots of non-equilibrium {Dirac} carrier
  distributions in graphene}.
\newblock \emph{\bibinfo{journal}{Nat. Mater.}} \textbf{\bibinfo{volume}{12}},
  \bibinfo{pages}{1119} (\bibinfo{year}{2013}).
\newblock \urlprefix\url{https://www.nature.com/articles/nmat3757}.

\bibitem{breusing2011ultrafast}
\bibinfo{author}{Breusing, M.} \emph{et~al.}
\newblock \bibinfo{title}{Ultrafast nonequilibrium carrier dynamics in a single
  graphene layer}.
\newblock \emph{\bibinfo{journal}{Phys. Rev. B}} \textbf{\bibinfo{volume}{83}},
  \bibinfo{pages}{153410} (\bibinfo{year}{2011}).
\newblock
  \urlprefix\url{https://journals.aps.org/prb/abstract/10.1103/PhysRevB.83.153410}.

\bibitem{brida2013ultrafast}
\bibinfo{author}{Brida, D.} \emph{et~al.}
\newblock \bibinfo{title}{Ultrafast collinear scattering and carrier
  multiplication in graphene}.
\newblock \emph{\bibinfo{journal}{Nat. Commun.}} \textbf{\bibinfo{volume}{4}},
  \bibinfo{pages}{1987} (\bibinfo{year}{2013}).
\newblock \urlprefix\url{https://www.nature.com/articles/ncomms2987}.

\bibitem{lindenberg2017visualization}
\bibinfo{author}{Lindenberg, A.~M.}, \bibinfo{author}{Johnson, S.~L.} \&
  \bibinfo{author}{Reis, D.~A.}
\newblock \bibinfo{title}{Visualization of atomic-scale motions in materials
  via femtosecond {{X}-ray} scattering techniques}.
\newblock \emph{\bibinfo{journal}{Annu. Rev. Mater. Res.}}
  \textbf{\bibinfo{volume}{47}}, \bibinfo{pages}{425--449}
  (\bibinfo{year}{2017}).
\newblock
  \urlprefix\url{https://www.annualreviews.org/doi/abs/10.1146/annurev-matsci-070616-124152}.

\bibitem{Ebi2}
\bibinfo{author}{Mohammed, O.~F.}, \bibinfo{author}{Yang, D.-S.},
  \bibinfo{author}{Pal, S.~K.} \& \bibinfo{author}{Zewail, A.~H.}
\newblock \bibinfo{title}{{4D} scanning ultrafast electron microscopy:
  Visualization of materials surface dynamics}.
\newblock \emph{\bibinfo{journal}{J. Am. Chem. Soc.}}
  \textbf{\bibinfo{volume}{133}}, \bibinfo{pages}{7708--7711}
  (\bibinfo{year}{2011}).
\newblock \urlprefix\url{http://dx.doi.org/10.1021/ja2031322}.

\bibitem{warren1990x}
\bibinfo{author}{Warren, B.~E.}
\newblock \emph{\bibinfo{title}{{{X}-ray Diffraction}}}
  (\bibinfo{publisher}{Dover Publications, New York}, \bibinfo{year}{1990}).

\bibitem{trigo2010imaging}
\bibinfo{author}{Trigo, M.} \emph{et~al.}
\newblock \bibinfo{title}{Imaging nonequilibrium atomic vibrations with {X}-ray
  diffuse scattering}.
\newblock \emph{\bibinfo{journal}{Phys. Rev. B}} \textbf{\bibinfo{volume}{82}},
  \bibinfo{pages}{235205} (\bibinfo{year}{2010}).
\newblock
  \urlprefix\url{https://journals.aps.org/prb/abstract/10.1103/PhysRevB.82.235205}.

\bibitem{holt1999determination}
\bibinfo{author}{Holt, M.} \emph{et~al.}
\newblock \bibinfo{title}{Determination of phonon dispersions from {X}-ray
  transmission scattering: The example of silicon}.
\newblock \emph{\bibinfo{journal}{Phys. Rev. Lett.}}
  \textbf{\bibinfo{volume}{83}}, \bibinfo{pages}{3317} (\bibinfo{year}{1999}).
\newblock
  \urlprefix\url{https://journals.aps.org/prl/abstract/10.1103/PhysRevLett.83.3317}.

\bibitem{stern2018mapping}
\bibinfo{author}{Stern, M.~J.} \emph{et~al.}
\newblock \bibinfo{title}{Mapping momentum-dependent electron-phonon coupling
  and nonequilibrium phonon dynamics with ultrafast electron diffuse
  scattering}.
\newblock \emph{\bibinfo{journal}{Phys. Rev. B}} \textbf{\bibinfo{volume}{97}},
  \bibinfo{pages}{165416} (\bibinfo{year}{2018}).
\newblock
  \urlprefix\url{https://journals.aps.org/prb/abstract/10.1103/PhysRevB.97.165416}.

\bibitem{rini2007control}
\bibinfo{author}{Rini, M.} \emph{et~al.}
\newblock \bibinfo{title}{Control of the electronic phase of a manganite by
  mode-selective vibrational excitation}.
\newblock \emph{\bibinfo{journal}{Nature}} \textbf{\bibinfo{volume}{449}},
  \bibinfo{pages}{72} (\bibinfo{year}{2007}).
\newblock \urlprefix\url{https://www.nature.com/articles/nature06119}.

\bibitem{graves2013nanoscale}
\bibinfo{author}{Graves, C.} \emph{et~al.}
\newblock \bibinfo{title}{Nanoscale spin reversal by non-local angular momentum
  transfer following ultrafast laser excitation in ferrimagnetic {GdFeCo}}.
\newblock \emph{\bibinfo{journal}{Nat. Mater.}} \textbf{\bibinfo{volume}{12}},
  \bibinfo{pages}{293} (\bibinfo{year}{2013}).
\newblock \urlprefix\url{https://www.nature.com/articles/nmat3597}.

\bibitem{fausti2011light}
\bibinfo{author}{Fausti, D.} \emph{et~al.}
\newblock \bibinfo{title}{Light-induced superconductivity in a stripe-ordered
  cuprate}.
\newblock \emph{\bibinfo{journal}{Science}} \textbf{\bibinfo{volume}{331}},
  \bibinfo{pages}{189--191} (\bibinfo{year}{2011}).
\newblock \urlprefix\url{https://science.sciencemag.org/content/331/6014/189}.

\bibitem{G-mobility-1}
\bibinfo{author}{Du, X.}, \bibinfo{author}{Skachko, I.},
  \bibinfo{author}{Barker, A.} \& \bibinfo{author}{Andrei, E.~Y.}
\newblock \bibinfo{title}{Approaching ballistic transport in suspended
  graphene}.
\newblock \emph{\bibinfo{journal}{Nat. Nanotech.}}
  \textbf{\bibinfo{volume}{3}}, \bibinfo{pages}{491--495}
  (\bibinfo{year}{2008}).
\newblock \urlprefix\url{https://www.nature.com/articles/nnano.2008.199}.

\bibitem{G-mobility-2}
\bibinfo{author}{Bolotin, K.~I.} \emph{et~al.}
\newblock \bibinfo{title}{Ultrahigh electron mobility in suspended graphene}.
\newblock \emph{\bibinfo{journal}{Solid State Commun.}}
  \textbf{\bibinfo{volume}{146}}, \bibinfo{pages}{351--355}
  (\bibinfo{year}{2008}).
\newblock
  \urlprefix\url{https://www.sciencedirect.com/science/article/pii/S0038109808001178}.

\bibitem{fugallo2014thermal}
\bibinfo{author}{Fugallo, G.} \emph{et~al.}
\newblock \bibinfo{title}{Thermal conductivity of graphene and graphite:
  collective excitations and mean free paths}.
\newblock \emph{\bibinfo{journal}{Nano Lett.}} \textbf{\bibinfo{volume}{14}},
  \bibinfo{pages}{6109--6114} (\bibinfo{year}{2014}).
\newblock \urlprefix\url{https://pubs.acs.org/doi/10.1021/nl502059f}.

\bibitem{mostofi2008wannier90}
\bibinfo{author}{Mostofi, A.~A.} \emph{et~al.}
\newblock \bibinfo{title}{wannier90: A tool for obtaining maximally-localised
  wannier functions}.
\newblock \emph{\bibinfo{journal}{Comput. Phys. Commun.}}
  \textbf{\bibinfo{volume}{178}}, \bibinfo{pages}{685--699}
  (\bibinfo{year}{2008}).
\newblock
  \urlprefix\url{https://www.sciencedirect.com/science/article/pii/S0010465507004936}.

\bibitem{marzari2012maximally}
\bibinfo{author}{Marzari, N.}, \bibinfo{author}{Mostofi, A.~A.},
  \bibinfo{author}{Yates, J.~R.}, \bibinfo{author}{Souza, I.} \&
  \bibinfo{author}{Vanderbilt, D.}
\newblock \bibinfo{title}{Maximally localized wannier functions: Theory and
  applications}.
\newblock \emph{\bibinfo{journal}{Rev. Mod. Phy.}}
  \textbf{\bibinfo{volume}{84}}, \bibinfo{pages}{1419} (\bibinfo{year}{2012}).
\newblock
  \urlprefix\url{https://journals.aps.org/rmp/abstract/10.1103/RevModPhys.84.1419}.

\bibitem{jung2013tight}
\bibinfo{author}{Jung, J.} \& \bibinfo{author}{MacDonald, A.~H.}
\newblock \bibinfo{title}{Tight-binding model for graphene $\pi$-bands from
  maximally localized wannier functions}.
\newblock \emph{\bibinfo{journal}{Phys. Rev. B}} \textbf{\bibinfo{volume}{87}},
  \bibinfo{pages}{195450} (\bibinfo{year}{2013}).
\newblock
  \urlprefix\url{https://journals.aps.org/prb/abstract/10.1103/PhysRevB.87.195450}.

\bibitem{baroni2001phonons}
\bibinfo{author}{Baroni, S.}, \bibinfo{author}{De~Gironcoli, S.},
  \bibinfo{author}{Dal~Corso, A.} \& \bibinfo{author}{Giannozzi, P.}
\newblock \bibinfo{title}{Phonons and related crystal properties from
  density-functional perturbation theory}.
\newblock \emph{\bibinfo{journal}{Rev. Mod. Phy.}}
  \textbf{\bibinfo{volume}{73}}, \bibinfo{pages}{515} (\bibinfo{year}{2001}).
\newblock
  \urlprefix\url{https://journals.aps.org/rmp/abstract/10.1103/RevModPhys.73.515}.

\bibitem{piscanec2004kohn}
\bibinfo{author}{Piscanec, S.}, \bibinfo{author}{Lazzeri, M.},
  \bibinfo{author}{Mauri, F.}, \bibinfo{author}{Ferrari, A.} \&
  \bibinfo{author}{Robertson, J.}
\newblock \bibinfo{title}{Kohn anomalies and electron-phonon interactions in
  graphite}.
\newblock \emph{\bibinfo{journal}{Phys. Rev. Lett.}}
  \textbf{\bibinfo{volume}{93}}, \bibinfo{pages}{185503}
  (\bibinfo{year}{2004}).
\newblock
  \urlprefix\url{https://journals.aps.org/prl/abstract/10.1103/PhysRevLett.93.185503}.

\bibitem{lazzeri2008impact}
\bibinfo{author}{Lazzeri, M.}, \bibinfo{author}{Attaccalite, C.},
  \bibinfo{author}{Wirtz, L.} \& \bibinfo{author}{Mauri, F.}
\newblock \bibinfo{title}{Impact of the electron-electron correlation on phonon
  dispersion: Failure of {LDA} and {GGA DFT} functionals in graphene and
  graphite}.
\newblock \emph{\bibinfo{journal}{Phys. Rev. B}} \textbf{\bibinfo{volume}{78}},
  \bibinfo{pages}{081406} (\bibinfo{year}{2008}).
\newblock
  \urlprefix\url{https://journals.aps.org/prb/abstract/10.1103/PhysRevB.78.081406}.

\bibitem{venezuela2011theory}
\bibinfo{author}{Venezuela, P.}, \bibinfo{author}{Lazzeri, M.} \&
  \bibinfo{author}{Mauri, F.}
\newblock \bibinfo{title}{Theory of double-resonant raman spectra in graphene:
  Intensity and line shape of defect-induced and two-phonon bands}.
\newblock \emph{\bibinfo{journal}{Phys. Rev. B}} \textbf{\bibinfo{volume}{84}},
  \bibinfo{pages}{035433} (\bibinfo{year}{2011}).
\newblock
  \urlprefix\url{https://journals.aps.org/prb/abstract/10.1103/PhysRevB.84.035433}.

\bibitem{hellman2011lattice}
\bibinfo{author}{Hellman, O.}, \bibinfo{author}{Abrikosov, I.} \&
  \bibinfo{author}{Simak, S.}
\newblock \bibinfo{title}{Lattice dynamics of anharmonic solids from first
  principles}.
\newblock \emph{\bibinfo{journal}{Phys. Rev. B}} \textbf{\bibinfo{volume}{84}},
  \bibinfo{pages}{180301} (\bibinfo{year}{2011}).
\newblock
  \urlprefix\url{https://journals.aps.org/prb/abstract/10.1103/PhysRevB.84.180301}.

\bibitem{hellman2013temperature}
\bibinfo{author}{Hellman, O.}, \bibinfo{author}{Steneteg, P.},
  \bibinfo{author}{Abrikosov, I.~A.} \& \bibinfo{author}{Simak, S.~I.}
\newblock \bibinfo{title}{Temperature dependent effective potential method for
  accurate free energy calculations of solids}.
\newblock \emph{\bibinfo{journal}{Phys. Rev. B}} \textbf{\bibinfo{volume}{87}},
  \bibinfo{pages}{104111} (\bibinfo{year}{2013}).
\newblock
  \urlprefix\url{https://journals.aps.org/prb/abstract/10.1103/PhysRevB.87.104111}.

\bibitem{hellman2013temperature2}
\bibinfo{author}{Hellman, O.} \& \bibinfo{author}{Abrikosov, I.~A.}
\newblock \bibinfo{title}{Temperature-dependent effective third-order
  interatomic force constants from first principles}.
\newblock \emph{\bibinfo{journal}{Phys. Rev. B}} \textbf{\bibinfo{volume}{88}},
  \bibinfo{pages}{144301} (\bibinfo{year}{2013}).
\newblock
  \urlprefix\url{https://journals.aps.org/prb/abstract/10.1103/PhysRevB.88.144301}.

\bibitem{debernardi1995anharmonic}
\bibinfo{author}{Debernardi, A.}, \bibinfo{author}{Baroni, S.} \&
  \bibinfo{author}{Molinari, E.}
\newblock \bibinfo{title}{Anharmonic phonon lifetimes in semiconductors from
  density-functional perturbation theory}.
\newblock \emph{\bibinfo{journal}{Phys. Rev. Lett.}}
  \textbf{\bibinfo{volume}{75}}, \bibinfo{pages}{1819} (\bibinfo{year}{1995}).
\newblock
  \urlprefix\url{https://journals.aps.org/prl/abstract/10.1103/PhysRevLett.75.1819}.

\end{thebibliography}

%
\vspace{20pt}
\noindent
{\bf \large Acknowledgments}
\begin{acknowledgments}
\noindent
The authors thank Jin-Jian Zhou for fruitful discussions. X.T. thanks the Resnick Sustainability Institute at the California Institute of Technology for fellowship support. This work was partially supported by the National Science Foundation under Grant No. DMR-1750613, which provided for theory development, and by the Department of Energy under Grant No. DE-SC0019166, which provided for numerical calculations and code development. This research used resources of the National Energy Research Scientific Computing Center, a DOE Office of Science User Facility supported by the Office of Science of the U.S. Department of Energy under Contract No. DE-AC02-05CH11231.
\\ 
\end{acknowledgments}

\noindent
{\bf \large Author Contributions}\\
M.B. conceived the research. X.T. developed the computational codes and carried out the
calculations. All authors analyzed the results and wrote the manuscript.

\vspace{20pt}
\noindent
{\bf \large Additional Information}\\
\hspace{20pt}
{\bf Supplementary Information} accompany this paper (at URL to be added by the editorial office).\\

\noindent
{\bf Competing financial interests:} The authors declare no competing financial interests.\\

\end{document}